\documentclass[12pt,preprint]{aastex}

\def\msun{{~M}_{\odot}}

\def\medd{{\dot M_{Edd}}}
\def\mdot{{\dot M}}
\def\be{\begin{equation}}
\def\ee{\end{equation}}

\begin{document}

\shorttitle{Outflow influences on ADAF} \shortauthors{F. G., Xie;
F., Yuan}

\title{The influences of outflow on the dynamics of inflow}

\author{Fu-Guo Xie,\altaffilmark{1,}\altaffilmark{2,}\altaffilmark{3}
 Feng Yuan\altaffilmark{1,}\altaffilmark{2}}

\altaffiltext{1}{Shanghai Astronomical Observatory, 80 Nandan Road,
Shanghai 200030, China} \altaffiltext{2}{Joint Institute for Galaxy
and Cosmology (JOINGC) of SHAO and USTC} \altaffiltext{3}{Graduate
School of the Chinese Academy of Sciences, Beijing 100039, China;
fgxie@shao.ac.cn}

\begin{abstract}

Both numerical simulations and observations indicate that in an
advection-dominated accretion flow most of the accretion material
supplied at the outer boundary will not reach the inner boundary.
Rather, they are lost via outflow. Previously, the influence of
outflow on the dynamics of inflow is taken into account only by
adopting a radius-dependent mass accretion rate $\dot{M}=\dot{M}_0
(r/r_{\rm out})^s$ with $s>0$. In this paper, based on a 1.5 dimensional
description to the accretion flow, we investigate this problem in
more detail by considering the interchange of mass, radial and
azimuthal momentum, and the energy between the outflow and inflow.
The physical quantities of the outflow is parameterized based on our
current understandings to the properties of outflow mainly from
numerical simulations of accretion flows. Our results indicate that
under reasonable assumptions to the properties of outflow, the main
influence of outflow has been properly included by adopting
$\dot{M}=\dot{M}_0 (r/r_{\rm out})^s$.

\end{abstract}


\section{Introduction}

There are now strong observational evidences for the existence of
outflow in accretion flow system. One of the best examples comes
from Sgr A*, the supermassive black hole located at our Galactic
center. The accretion flow in this source is likely in the form of
the advection-dominated accretion flow (ADAF, or radiatively
inefficient accretion flow; Yuan, Quataert \& Narayan 2003). On one
hand, from the observational results from $Chandra$ combined with
the Bondi accretion theory we can calculate the value of the mass
accretion rate at the outer boundary---the Bondi radius. On the
other hand, radio polarization observations constrain the accretion
rate at the innermost region of the accretion flow nearly two orders
of magnitude lower than that determined at the Bondi radius (e.g.,
Marrone et al. 2006). This implies that about $99\%$ of the material
available at the Bondi radius will not finally enter into the black
hole horizon, rather, they must be lost in the form of outflow.
Outflow seems to exist also in more luminous sources whose accretion
mode is different from the ADAF. For example, the blueshifted
absorption lines, which indicates the existence of outflowing
materials, have been detected in the X-ray spectrum of some Seyfert
1 sources (e.g., NGC 3783: Kaspi et al. 2001) and more spectacularly
in quasars (e.g., PG 1115+80: Chartas, Brandt \& Gallagher 2003).
The existence of outflow has been paid more and more attention
recently in the field of galaxy formation because of its feedback
effect in the coevolution of galaxy and the central active galactic
nuclei (e.g., Silk \& Rees 1998; Granato et al. 2004; Springel et
al. 2005).

Many work has been done on the origin and dynamics of outflow
(e.g., Xu \& Chen 1997; Blandford \& Begelman 2004;
Xue \& Wang 2005) in the frame of self-similar hydrodynamical
solution. Magnetic field, especially its
poloidal component, may presumably serve as the most promising
mechanism on producing outflow, as proposed by, e.g., Blandford \&
Payne (1982). This has been confirmed in non-radiative
magnetohydrodynamical (MHD) numerical simulations of accretion flows
(e.g., Stone \& Pringle 2001; Igumenshchev et al. 2003; Vlahakis \&
K$\ddot{o}$nigl 2003; McKinney 2006). Radiation pressure could be
another important mechanism in luminous accretion disk (Proga 2003).
But even in the absence of magnetic field and strong radiation,
outflow is likely present in ADAFs. This is first proposed
from analytical argument that the Bernoulli parameter of an ADAF
is large or even positive because of the small radiative energy
loss. This implies that the gas is inclined to escape once they are
perturbed (Narayan \& Yi 1994; Blandford \&
Begelman 1999). This suggestion was later confirmed by numerical
simulation by Stone, Pringle \& Begelman (1999).

Since the outflow is likely very strong (Misra \& Taam 2001), they
may provide an additional important sink of angular momentum and energy.
So it is important to investigate its dynamical influence to
inflow. This problem has been investigated by Kuncic \& Bicknell (2007)
in the context of the standard thin disk. In this paper we focus on ADAFs.
Blandford \& Begelman (1999, hereafter BB99) examined this
question through a one-dimensional self-similar approach. A
phenomenological way was adopted in which they parameterized the
rate at which mass, angular momentum and energy are extracted
through outflow, regardless the mechanism of the formation of
outflow.

BB99 gives a quite general description, covering a broad kind of
outflow including Poynting flux whose mass flux is zero while energy
flux is not. It is based on self-similar assumption. For the purpose
of application and comparison with observation, however, we need to
discuss it based on global solutions, because self-similar solution
is too simplified to be used to calculate the emitted spectrum.
Quataert \& Narayan (1999) presented the first effort on this
aspect. They calculate the global solution of inflow when strong
outflow is present by using a radius-dependent mass accretion rate,
$\dot{M}\propto r^s$ while keeping all other equations describing
inflow such as the momentum and energy equations unchanged. This is
roughly equivalent to assuming that the specific angular momentum
and energy of outflow is identical to the inflow at the same radius
where outflow is launched (see e.g., eqs.(\ref{mass})-(\ref{ion})).
This approach is subsequently adopted in almost all following works
(e.g., Yuan, Quataert \& Narayan 2003). In this paper, we refer to
this treatment as standard treatment.

 While the approximation of Quataert
\& Narayan (1999; see also Yuan, Quataert \& Narayan 2003) may
capture the most important influence of outflow to inflow, it is not
obvious in what degree we can use this approximation or how good
this approximation is when we compare the theoretical prediction
such as the spectrum to observations. This is the aim of the present
paper. More specifically, by considering the conservations of fluxes
of mass, momentum, and energy of the combined inflow/outflow system,
we focus on the influence of outflow on the dynamics of inflow. We
will use a ``1.5 dimension'' description of the accretion flow,
which means the height-integrated equations will be used instead of
a fully two-dimensional description, but the conservation equations
take into account the outflow in the vertical direction as well. The
paper is organized as follows: in \S 2 we present basic equations
for our model and discuss the main properties of outflow. The
calculation results are presented in \S 3. The last section is
devoted to a summary.

\section{Accretion Model with Outflow }

\subsection{Basic Equations}

\begin{figure}
\centering
\includegraphics[width=0.9\columnwidth,clip=true]{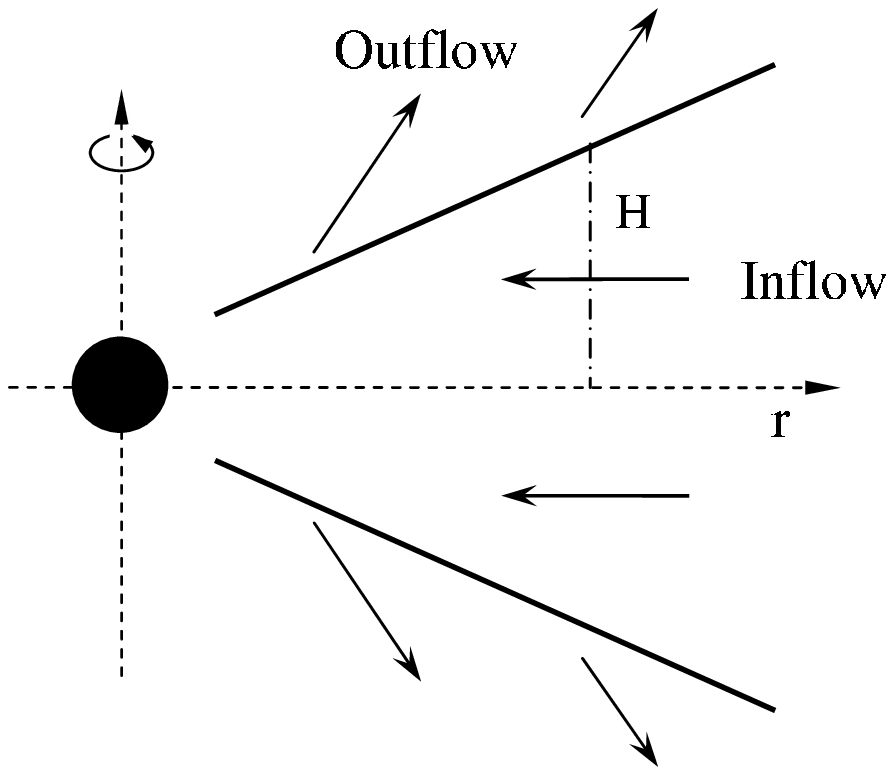}
\caption{Schematic diagram of inflow/outflow model. Outflow launches
from the surface $z = H$.} \vspace{0.1in}\label{scheme}
\end{figure}

We adopt a cylinder coordinate ($r, \phi, z$) to describe a steady
axisymmetric ($\partial /\partial t= \partial / \partial\phi = 0 $)
accretion flow. The Paczy\'nski \& Wiita potential (Paczy\'nski \&
Wiita 1980) $\psi = -GM_{BH}/(r-r_g)$ is adopted to mimic the
geometry of a Schwarzschild black hole, with $M_{BH}$ is the mass of
black hole and $r_g \equiv 2 G M_{BH}/c^2$ is the Schwarzschild
radius of the black hole. As shown in Fig. \ref{scheme}, we divide
the whole accretion flow at each radius into two parts, i.e., inflow
and outflow. For the inflow, we assume a hydrostatic balance in
vertical direction ($\upsilon_z = 0$ for the inflow) and assume all
quantities such as the radial and azimuthal velocity ($\upsilon_r$
and $\upsilon_\phi$), ions and electron temperature ($T_i, T_e$) and
the sound speed ($c_s$) are only functions of radius $r$. Such an
isothermal assumption in the vertical direction results in a density
distribution of $\rho(r,z) = \rho(r,0) \exp(-z^2/2 H^2)$ in the
inflow, where $H = c_s/\Omega_K$ is the vertical scale height of
inflow\footnote{This kind of vertical density structure is based on
a first-order expand of Paczy\'nski \& Wiita potential, which is not
exact far away from the equatorial plane (Gu \& Lu 2007).}. We set
$z=H$ as the surface where an outflow launches. The vertical
gradients of above quantities are absorbed by their discontinuity
between inflow and outflow at this surface, except that the density
distribution is continuous and the density of outflow at $z=H$ is
then $e^{-1/2} \rho(r,0)$. Note that the vertical velocity of
outflow $\upsilon_{z,w}$, will ``compress'' the inflow because of
momentum conservation, thus the vertical scale height may be
smaller. We neglect this effect here.

From compressible Navier-Stokes equations, we can write the
equations of the conservations of mass, momentum, and energy for the inflow as
follows (see Appendix A for details):
\begin{equation}
\label{mass} \frac{d \mdot(r)}{d r} = \eta_1 4 \pi r \rho
\upsilon_{z,w}
\end{equation}
\begin{equation}
\label{radial} \upsilon_r \frac{d \upsilon_r}{d r} + \eta_1
\upsilon_{z,w} \frac{\upsilon_{r,w}-\upsilon_r}{H} = r
(\Omega^2-\Omega_k^2) -\frac{1}{\rho}\frac{d P}{d r} -\frac{1}{2}
\frac{d c_s^2}{d r}
\end{equation}
\begin{equation}
\label{angular} \rho r \upsilon_r  \frac{d}{d r}(r^2 \Omega)+ \eta_1
r^2 \rho \upsilon_{z,w}
\frac{\upsilon_{\phi,w}-\upsilon_\phi}{H}=\frac{1}{H}\frac{d}{d r}
(r^2 H \tau_{\phi r})
\end{equation}
\begin{equation}
\label{electron} \rho \upsilon_r(\frac{d \epsilon_e}{d
r}-\frac{p_e}{\rho^2}\frac{d \rho}{d r}) + \eta_1 \rho
\upsilon_{z,w} \frac{\epsilon_{e,w}-\epsilon_e}{H}= \delta q^+
+q_{ie} - q^-
\end{equation}
\begin{equation}
\label{ion} \rho \upsilon_r(\frac{d \epsilon_i}{d
r}-\frac{p_i}{\rho^2}\frac{d \rho}{d r}) + \eta_1 \rho
\upsilon_{z,w} \frac{\epsilon_{i,w}-\epsilon_i}{H}= (1-\delta) q^+ -
q_{ie}
\end{equation}
Here all quantities have their usual meanings. The specific internal
energy of electrons and ions are $\epsilon_e, \epsilon_i$,
respectively. The pressure $P$ is the sum of gas and magnetic
pressure $P=P_{gas} + P_{mag}$. The inflow's accretion rate is
defined as $\dot M(r) \equiv -4 \pi r \rho \upsilon_r H$ and $\eta_1
\equiv \rho_w/\overline{\rho}$ is the density ratio of outflow and
inflow (Appendix A). Parameter $\delta$ describes the fraction of
the turbulent energy dissipation rate $q^+$ ($\equiv \tau_{\phi r} r
d \Omega/d r$) that heats electrons directly. Energy transfers from
ions to electrons through Coulomb collisions at a volume rate
$q_{ie}$, and radiative cooling rate is denoted by $q^-$. The
quantities with subscript {\it w} denote the quantities of
wind/outflow just away from the launching surface $z = H$. We take
the $\alpha$ viscosity description for the stress tensor $\tau_{\phi
r}$ (Shakura \& Sunyaev 1973):
\begin{equation}
\label{vis} \tau_{\phi r} = - \alpha P
\end{equation}
where $\alpha$ is the dimensionless viscosity parameter. Other
stress tensor components are neglected for simplicity, except that
the $\phi z$ component is considered by taking into account the
angular momentum exchange between inflow and outflow at $z=H$.

Obviously, it is impossible to directly solve the eqs.
(\ref{mass})-(\ref{ion}). We therefore introduce the following
parameters ``$\xi$'' to evaluate the radial, azimuthal, and vertical
velocity, and the ion and electron temperatures of outflow in terms
of inflow, \be \upsilon_{r,w}=\xi_r \upsilon_{ff},\ee \be
\upsilon_{\phi,w}=\xi_\phi \upsilon_{\phi}, \ee \be
\upsilon_{z,w}=\xi_z c_s,\ee \be T_{i,w}=\xi_{T_i} T_i, \ee \be
T_{e,w}=\xi_{T_e} T_e.\ee Here $\upsilon_{ff}$ are the free-fall
velocity, $\upsilon_{\phi}$, and $c_s$ are azimuthal velocity and
the sound speed of the inflow, respectively. We assume that these
parameters are independent of radius. While this assumption is
simple, we think it can capture the main physics of the influence of
outflow in a reasonable way. Specifically, this simple assumption
does not mean all quantities are a power-law function of radius as
the usual ``power-law'' assumption  of the mass flux of inflow. If
we know the values of these parameters, we will be able to get the
global solution of eqs. (\ref{mass})-(\ref{ion}).

\subsection{Outflow's Properties}

We now estimate the properties of outflow. Generally all these
quantities should be a function of $z$. Here we consider the
properties of outflow when they are just launched or detached from inflow. All
their subsequent evolution should be due to outflow itself and does
not affect the inflow any longer.

The first quantity is the strength of the outflow, or the mass lost
rate. BB99 assume $\mdot \propto r^{s}$ with $0 \leq s < 1$. This
ensures that the mass accretion rate decreases while the released
energy increases with accretion (BB99). The strength of outflow in
our notion is mainly governed by $\xi_z$. The above range
corresponds to $0 \leq \xi_z < - \upsilon_r/\eta_1 c_s \approx 0.2$
(ref. eq. (\ref{ratio}), but note $s(r)$ now is a function of $r$).

We next consider the value of $\xi_{\phi}$. The vertical
distribution of angular momentum of the accretion flow is
complicated. Two-dimensional self-similar analysis on ADAF based on
hydrodynamics shows that the specific angular momentum of outflow is
lower than inflow (Narayan \& Yi 1995; Xu \& Chen 1997; Blandford \&
Begelman 2004). This result is confirmed later by numerical
simulations (e.g., Stone, Pringle \& Begelman 1999). However, any
magnetic coupling between inflow and outflow will likely lead to
transportation of angular momentum from the former to latter (Spruit
1996; Stone \& Pringle 2001; BB99; Blandford \& Begelman 2004).
Therefore in this paper we explore $\xi_{\phi}$ in a range around
unity, $0.8<\xi_{\phi}<1.2$.

Hydrodynamical and MHD simulations also reveal that the specific
internal energy or the temperature of the gas increases from the
equator to higher altitude (e.g., Stone, Pringle \& Begelman 1999;
De Villiers et al. 2005; Beckwith, Hawley \& Krolik 2008).
One underlying reason may be that the gas with higher internal
energy may escape more easily. We therefore consider $\xi_{T_i}=1,
1.5$ and $\xi_{T_e} =1, 1.5$.

 It is highly unclear about the radial velocity of outflow when they
are just launched although we somehow know how they will be
accelerated later. But we speculate that it should be positive, and
should not be larger than a fraction of the local Keplerian or
free-fall velocity $\upsilon_{ff}$. Fortunately, although it
may be important for the dynamics of outflow itself, we find that
the value of $\xi_r$ has minor effect on the inflow. We simply set
$\xi_r \equiv 0.2$ in our calculations.

\section{Results}

\begin{figure}
\epsscale{1.} \plotone{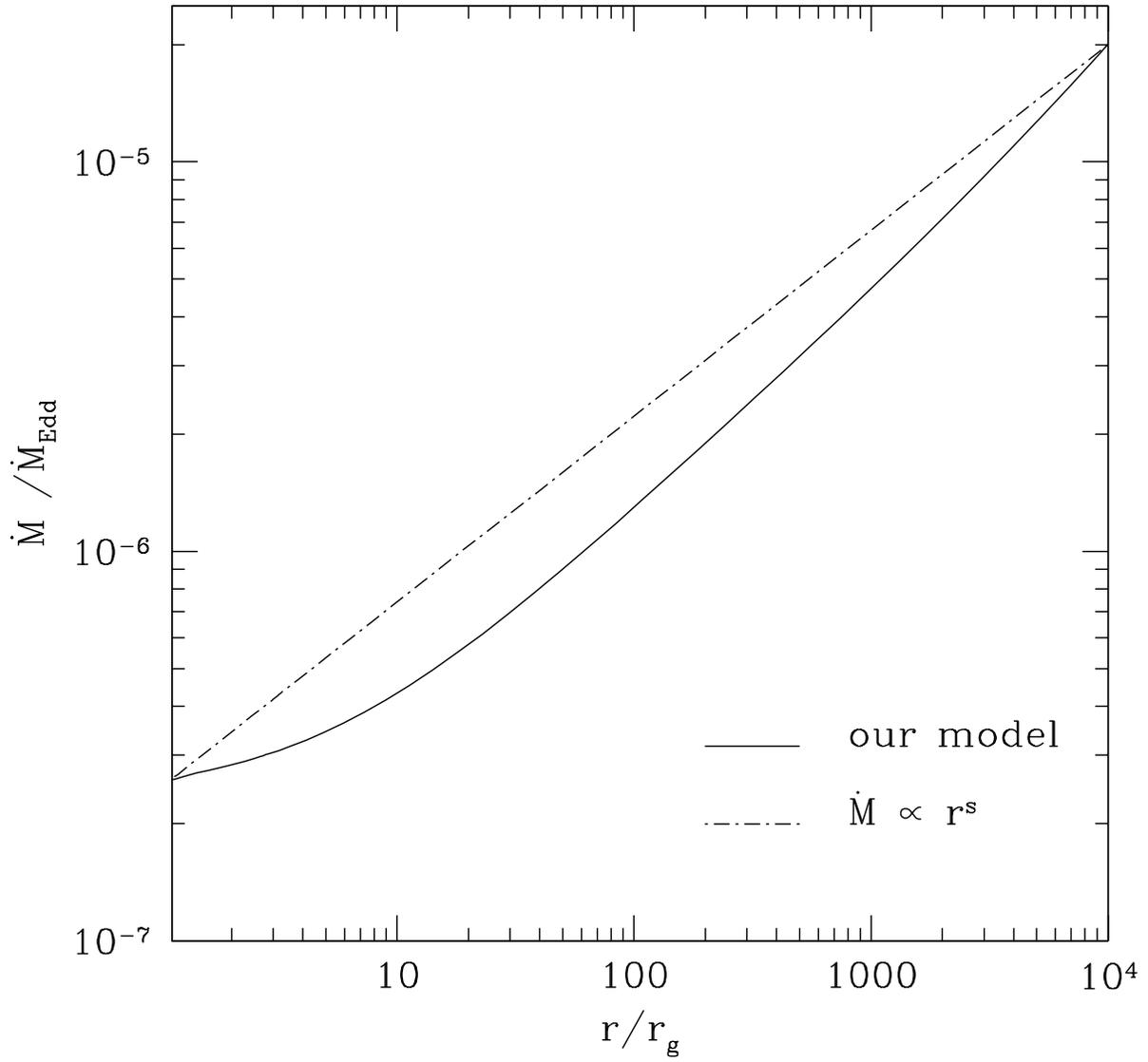}
\caption{The change of accretion rate as a function of radius for
our model (solid line) and the standard treatment (dashed line) in
which $\dot{M}(r)\propto r^s$ with $s$ being a constant.}
\vspace{0.1in}\label{mdot}
\end{figure}

For our specific model, we adopt the black hole mass $M_{BH} \equiv
4.0\times 10^6 \msun$, accretion rate at the outer boundary $r_{out}
= 10^4 r_s$ is $\mdot=1.1 \times 10^{-5} \medd$, where $\medd = {\rm
10 L_{Edd}}/c^2$, is the Eddington accretion rate. The values for
other parameters are $\alpha = 0.1, \beta \equiv P_{gas}/P_{total} =
0.9, \delta = 0.3$. These parameters are close to those in Yuan,
Quataert \& Narayan (2003) to model the supermassive black hole in
our Galactic center. There they assume $\mdot\propto r^s$ with
$s=0.27$ being a constant. Under our notation, we have
\begin{equation}
\label{ratio} s(r) = \frac{d \ln \mdot(r)}{d \ln r} = \eta_1 \xi_z
\frac{\upsilon_k}{- \upsilon_r},
\end{equation}
where $\upsilon_k$ is the Keplerian velocity. We would like to note
that $s(r)$ now is not a constant as in Yuan, Quataert \& Narayan (2003)
(or Quataert \& Narayan 1999). The slope of $\mdot(r)$ now
is steeper at large radius while flatter at small radius, because of
the quicker increase of $v_r$ compared to $\upsilon_k$. This is
shown in Fig. \ref{mdot}, where we adjust the parameter $\xi_z$ so
that the accretion rates at $r_{out}$ and horizon are the same as
the case of $\mdot\propto r^s$ with $s=0.48$.

\begin{figure*} \epsscale{1.} \plotone{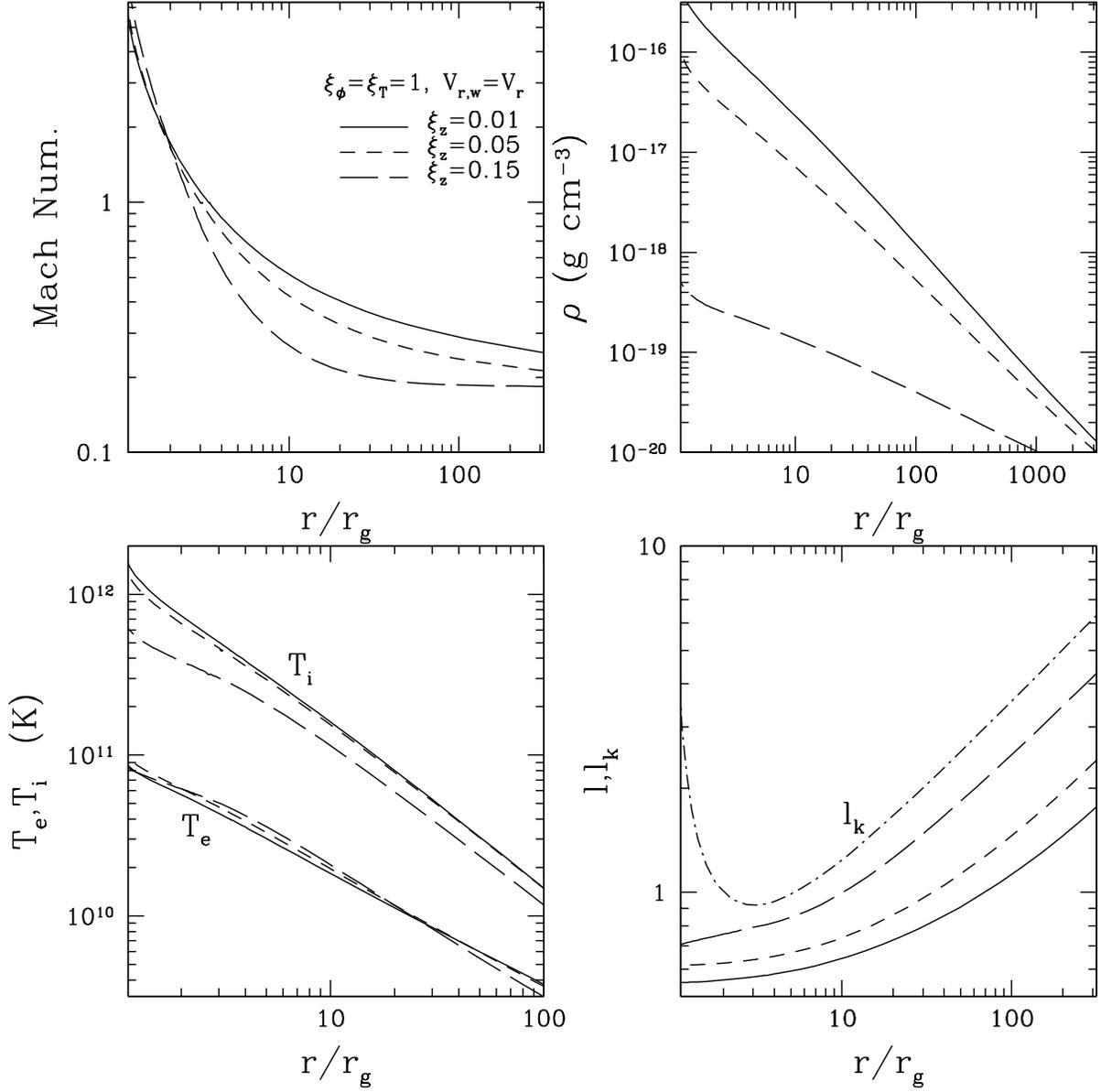}
\caption{Influence of the vertical velocity of the outflow described
by $\xi_z$ on the Mach number, density, temperature, and specific
angular momentum of inflow. The solid, dashed, and long-dashed lines
are for $\xi_z=0.01, 0.05, 0.15$, respectively. The Keplerian
angular momentum $l_k$ is show as the dot-dashed line in the bottom
right panel. Other parameters are $\upsilon_{r,w} = \upsilon_r,
\xi_\phi= \xi_{T_e}= \xi_{T_i} = 1.0$.} \vspace{0.1in} \label{xiz}
\end{figure*}

We first investigate the effect of $\xi_z$. Fig. \ref{xiz} shows the
effects of various $\xi_z$ on inflow, with other outflow parameters
fixed at $\xi_\phi = \xi_{T_e}= \xi_{T_i} = 1.0$ and $\upsilon_{r,w}
= \upsilon_r$. The four plots show the Mach number, profiles of
density, temperature, and specific angular momentum. The dotted,
dashed, and long-dashed lines correspond to $\xi_z = 0.01, 0.05$ and
$0.15$, respectively. As the outflow becomes stronger ($\xi_z$
increases), the gas density decreases while the ion temperature
decreases. This is very similar to the case of the standard
treatment (with increasing $s$). The decrease of ions temperature is
because when more and more accretion material is lost via the
outflow, the density profile becomes flatter thus the compression
work which is an important heating mechanism for ions becomes
weaker. Different from the ions, the electrons temperature has no
obvious relation with the strength of outflow. This is because
different from ions the compression work in the electron energy
equation is about one order of magnitude smaller due to the lower
electron temperature (ref. eq. \ref{electron}).

We mentioned before that the value of the radial velocity of outflow
or equivalently $\xi_r$, has minor effect on the dynamics of inflow.
The ``kick back'' force due to the discrepancy of the radial
velocity between the inflow and outflow is manifested by the second
term in eq. (\ref{radial}). Since $\upsilon_{z,w} = \xi_z c_s,
H=c_s/\Omega_k$ and $\upsilon_r \sim \alpha \upsilon_k$, this term
is roughly $\alpha \xi_z (\ll 1)$ times of the gravitational force
thus can be neglected. So we simply fix $\xi_r = 0.2$ in this paper.

We now check how good the standard treatment is. For this purpose,
we first get the global solution with the standard treatment with
$\mdot=2\times 10^{-5} (r/r_{out})^{0.25}$ and $r_{out}=10^4r_g$.
We then get the global solution of eqs. (\ref{mass}) - (\ref{ion}) for
various sets of outflow parameters of $\xi_{\phi}=0.8, 1.0, 1.2$,
$\xi_{T_i}=1.0$ (referred to Case A) and $1.5$ (referred to as Case
B), and $\xi_{T_e}=1.5$. For each set of these parameters, we adjust
the value of $\xi_z$ so that the mass accretion rates at $r_{out}$
and black hole horizon are equal to the values in the above standard
treatment. By doing this, we want to focus on the influence on inflow of
the transportation of angular momentum and internal energy between
inflow and outflow, which is neglected in the standard treatment. Note the
profile of accretion rate in this case is similar to Fig. 2.

\begin{figure*} \epsscale{1.} \plotone{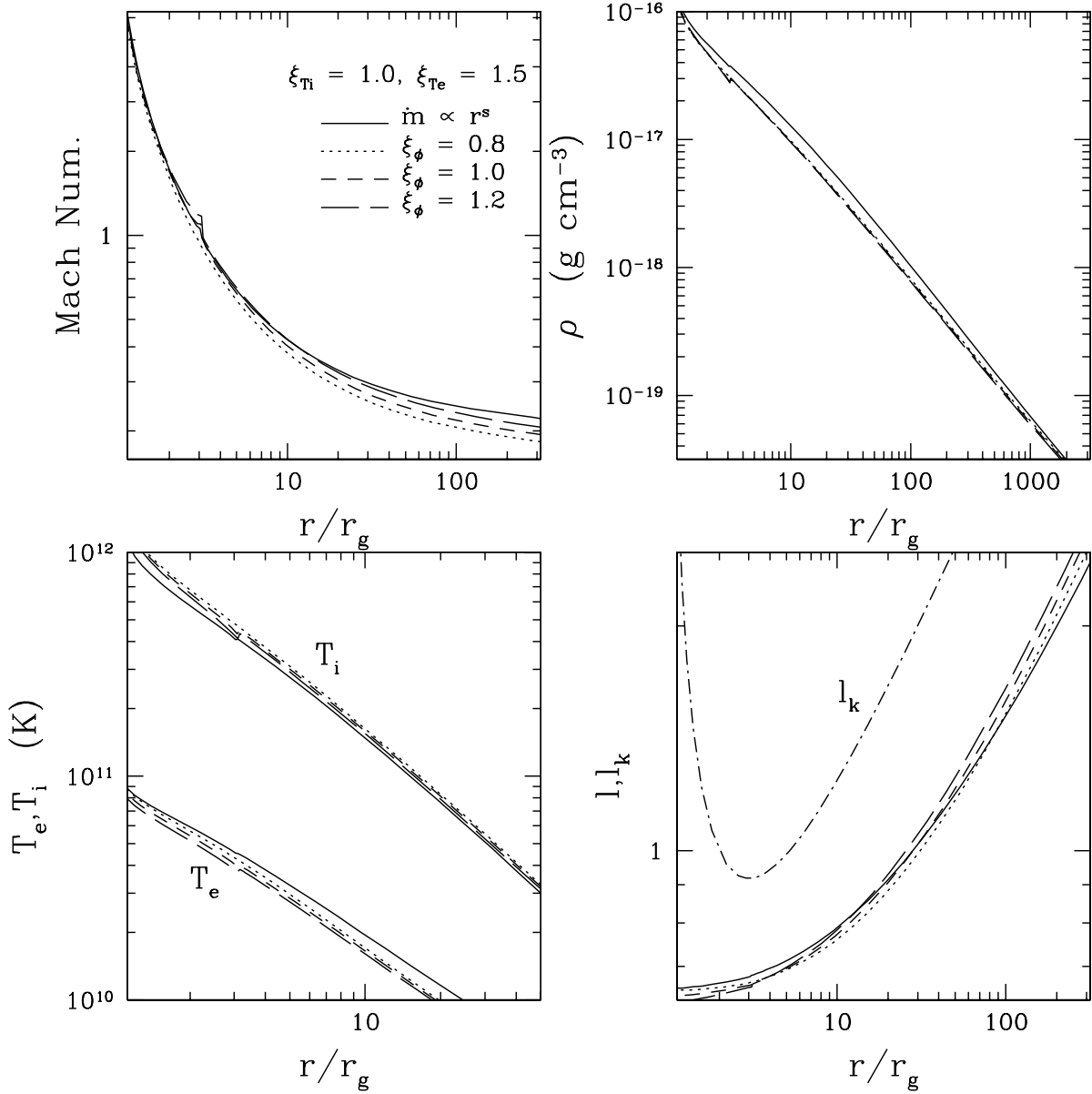}
\caption{Influence of the specific angular momentum of the outflow
described by $\xi_{\phi}$ on the dynamics of inflow and their
comparison with the standard treatment (solid line). The dotted,
dashed, and long-dashed lines are for $\xi_{\phi}=0.8, 1.0, 1.2$,
respectively. Other parameters are $\xi_{T_i}=1$ (Case A),
$\xi_r=0.2, \xi_{T_e}=1.5$. We adjust $\xi_z$ so that all the four
models have the same accretion rates at the inner and outer
boundary.} \vspace{0.1in}\label{mratedyn1}
\end{figure*}

Fig. \ref{mratedyn1} shows the comparison of Case A with the
standard treatment. We can see that our models intend to have lower
densities compared to the standard treatment, although the accretion
rates at the outer and inner boundaries are the same. We can easily
understand this by looking at bottom-left panel of Fig.
\ref{mratedyn1}. Our solutions have higher ion temperature and lower
electron temperature at the inner region of the inflow. The higher
ion temperature is because the density profile in the inner region
is steeper thus the compression heating is stronger. The lower
electron temperature is because $\xi_{T_e}=1.5>1$ which implies that
some internal energy is transferred into the outflow from inflow
(ref. eq. \ref{electron}).

\begin{figure*} \epsscale{1.} \plotone{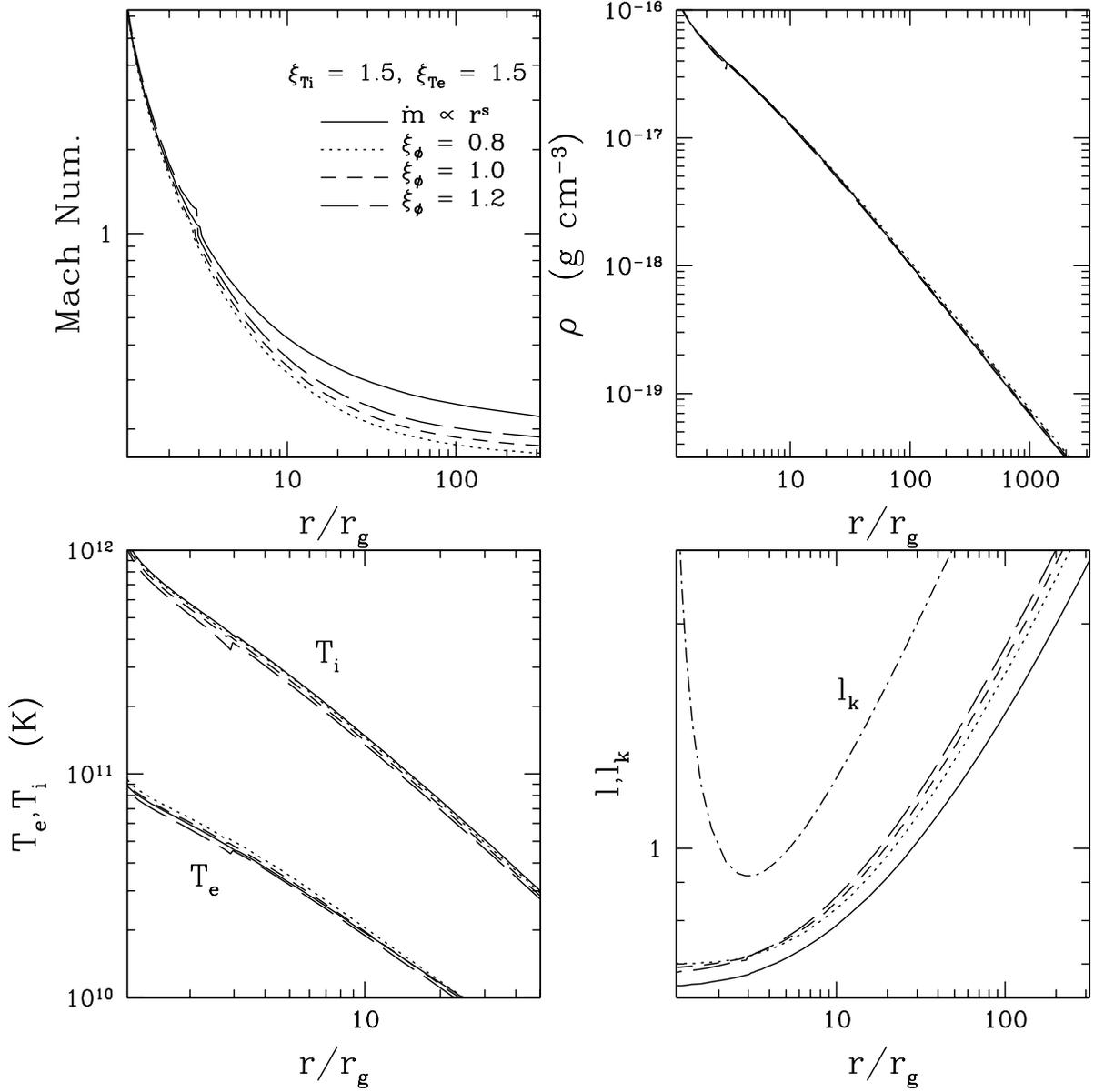}
\caption{Influence of the specific angular momentum of the outflow
on the dynamics of inflow and their comparison with the standard
treatment. All parameters are the same as Fig. \ref{mratedyn1} (Case
A) except $\xi_{T_i}=1.5$.} \vspace{0.1in}\label{mratedyn2}
\end{figure*}

Fig. \ref{mratedyn2} shows the dynamical influences of outflow for
Case B. Compared to Case A, the ion temperature is lower. This is
obviously because $\xi_{T_i}$ is larger, $\xi_{T_i}=1.5$, so some
internal energy is transferred from inflow to outflow. The lower ion
temperature results in a smaller $H$. This, combined with the
smaller radial velocity, make the density of inflow higher compared
to Case A and almost identical to that of the standard treatment, as
shown in the figure.
We also see from
the figure that both the value of the specific angular momentum and
its slope are higher compared to the standard treatment. This
results in a stronger viscous heating, which somehow cancel the
effect of $\xi_{T_e}=1.5>1$. This is why the electron temperature is
roughly the same with the standard treatment while higher than that
of Case A.

From Fig. \ref{xiz} to \ref{mratedyn2}, we find that within the
range of the values of parameters we adopt to describe the
outflow, the strength of the outflow has the most significant influence
on the dynamics of inflow (i.e., Fig. \ref{xiz}).
The influences of all other properties of outflow, namely
the angular momentum, temperature,
and radial velocity, are much smaller (Figs. \ref{mratedyn1}
and \ref{mratedyn2}). This is the reason why the discrepancy
between our model (with different properties of outflow) and
the standard treatment, which only considers the strength of outflow but
assume the properties of outflow are the same with inflow, is small.

\section{Summary}

Outflow is now believed to be very significant in
advection-dominated accretion flow thus it is important to
investigate their influence on the dynamics of inflow. Previously
this was done by using a rather simple way. In the ``standard
treatment'' the only change compared to the case of no outflow is
that the mass accretion rate is not a constant, but a power-law
function of radius, $\dot{M}\propto r^s$ ($s>0$). All other
equations describing the accretion flow remain unchanged (e.g.,
Quataert \& Narayan 1999; Yuan, Quataert \& Narayan 2003).

In this paper, we investigate the influence of outflow in more
detail to check how good the above ``standard treatment'' is. We
have derived the height-integrated accretion equations including the
coupling between the inflow and outflow, to investigate the
influence of outflow on the dynamics of inflow. We assume
hydrostatic equilibrium for the inflow. For the outflow, we assume
they are launched just above the surface of the inflow. We
parameterize and estimate the quantities of outflow in terms of the
quantities of inflow mainly from the results of numerical
simulations. In this way, we reduce the numbers of unknown
quantities in the above inflow/outflow equations thus we are able to
get their global solution.

We have studied the influences on the dynamics of inflow of the
strength (via vertical velocity), the (ion and electron)
temperature, specific angular momentum, and the radial velocity of
the outflow. We find that among them the strength of the outflow is
the most important quantity. It can produce orders of magnitude
difference for the density of inflow. If the strength of outflow is
fixed, all other quantities of outflow can only produce a difference
for the density and temperature within a factor of $\sim$ two, if
our estimations to the properties of outflow are correct. Therefore,
the ``standard treatment'' is usually a good approximation.

The largest uncertainty in our model comes from the
estimations to the properties of outflow such as their temperature,
specific angular momentum, azimuthal and radial velocity. We
estimate these values from numerical simulations to accretion flows
which is still not exact. With the rapid development of numerical
simulation, our results will be significantly improved. Especially,
if the properties of outflow is found to be far more deviated
from the inflow than those adopted in the present paper
(e.g., $\xi_{T_e} >> 1$), the standard treatment will be not enough
although we believe that our conclusion that the strength of
outflow is the most influential quantity should remain correct. In
that case, the influence of the other properties of outflow
must be taken into account as well.

\acknowledgments We thank our referee, Chris Done,
for constructive suggestions. This work was supported in part by the
Natural Science Foundation of China (grant 10773024),
One-Hundred-Talent Program of China, and Shanghai Pujiang Program.

\begin{appendix}

\section{The height-integrated equations describing inflow/outflow}


For a steady axisymmetric accretion flow, the equation of mass conservation is:
\begin{equation}
\label{cont} \frac{1}{r}\frac{\partial}{\partial r} (r \rho
\upsilon_r) + \frac{\partial}{\partial z} ( \rho \upsilon_z) = 0.
\end{equation}
Defining the mass accretion rate $\mdot \equiv -4 \pi r \rho
\upsilon_r H$, we integrate eq. (\ref{cont}) in $z$ from $0$ to
$H^+$, here $H^+$ denotes just above the surface $z=H$ where the
outflow is just launched. Noting $\upsilon_z=0$ for inflow and
$\upsilon_z=\upsilon_{z,w}$ for outflow at $H^+$, we get:
\begin{equation}
\frac{d \mdot}{d r} = \eta_1 4 \pi r \rho \upsilon_{z,w},
\end{equation}
where \be \eta_1 = \frac{\rho_w}{\overline{\rho}} =
\frac{e^{-1/2}\rho(r,0)}{\frac{1}{H} \int_0^H \rho(r,0)
exp(-\frac{z^2}{2 H^2}) d z} = 0.7089, \ee which gives the ratio of
the density of outflow and height-averaged inflow.

The radial and azimuthal momentum equations read as follows:
\begin{equation}
\label{rmom} \rho (\upsilon_r \frac{\partial \upsilon_r}{\partial r}
- \frac{\upsilon_\phi^2}{r} + \upsilon_z \frac{\partial
\upsilon_r}{\partial z}) = -\frac{\partial P}{\partial r} + \rho
g_r,
\end{equation}
\begin{equation}
\label{phimom} \rho (\upsilon_r \frac{\partial
\upsilon_\phi}{\partial r} + \frac{\upsilon_r \upsilon_\phi}{r} +
\upsilon_z \frac{\partial \upsilon_\phi}{\partial z}) =
\frac{1}{r^2}\frac{\partial}{\partial r}(r^2 \tau_{\phi r}).
\end{equation}
Here $g_r$ is the radial component of the gravitational force.
The energy equation is
\begin{equation}
\rho Tds/dt\equiv \rho\left(dU/dt - P/\rho^2 d \rho/ d
t\right) = q^+ - q^-,
\end{equation}
which for steady flow reduces to:
\begin{equation}
\label{energy} \rho \left[ \upsilon_r \left(\frac{\partial
U}{\partial r} - \frac{P}{\rho^2} \frac{\partial \rho}{\partial
r}\right) + \upsilon_z \left(\frac{\partial U}{\partial z} -
\frac{P}{\rho^2} \frac{\partial \rho}{\partial z}\right)\right]=q^+
- q^-.
\end{equation}

We integrate the above equations \ref{rmom}-\ref{energy} for $z$
from $z=0$ to $z=H^+$ using the following general result:
\begin{equation}
\label{hint}\int_0^{H^+} f \upsilon_z \frac{\partial g}{\partial z}
d z = f \upsilon_{z,w} (g_w - g_{z=H}),
\end{equation}
where $f$ and $g$ are functions of ($r, z$). Note that if $g(r,z)$
is continuous at $z = H$ (e.g., density $\rho$), the right side of
equation (\ref{hint}) is equal to 0. Then we will get eqs.
(\ref{radial}) - (\ref{ion}).

\end{appendix}

{}


\end{document}